\documentclass[aps,pr,twocolumn,floatfix]{revtex4}
\usepackage{amsmath,graphicx,xspace,units}

\graphicspath{{figures/}}

\renewcommand{\vec}[1]{\mathbf{#1}}
\newcommand{\vecr}{\ensuremath{{\vec r}}\xspace}
\newcommand{\ymax}{\ensuremath{y_\text{max}}\xspace}

\renewcommand{\Im}{\mathrm{Im}}

\newcommand{\abs}[1]{\lvert #1 \rvert}

\begin{document}

\title{Transport and Fractionation in Periodic Potential-Energy Landscapes}

\author{Kosta Ladavac}
\author{Matthew Pelton}

\affiliation{Dept.\ of Physics and James Franck Institute\\
The University of Chicago, Chicago, IL 60637}

\author{David G. Grier}

\affiliation{Dept.\ of Physics and Center for Soft Matter Research\\
New York University, New York, NY 10003}

\date{\today}

\begin{abstract}
Objects driven through periodically modulated potential-energy
landscapes in two dimensions
can become locked in to symmetry-selected directions 
that are independent of
the driving force's orientation.
We investigate this problem in the overdamped limit, and
demonstrate that the crossover from free-flowing to locked-in
transport can depend exponentially on an object's size,
with this exceptional selectivity emerging
from the periodicity of the environment.
\end{abstract}


\maketitle

\section{Introduction}
The theme of transport through modulated potential-energy landscapes
pervades solid-state physics 
and arises in many natural
and industrial processes.
This problem has been studied extensively in the
quantum-mechanical limit.
Considerably less attention has been paid to
the classical limit, where effects such as viscous damping
and thermal randomization complicate the analysis.
This article focuses on noninertial transport of classical
objects driven through periodically modulated potential-energy landscapes
by constant, uniform forces.
The one-dimensional variant of this problem 
has been thoroughly investigated \cite{risken89}, and its
results have been applied profitably
to such processes as
gel electrophoresis.
We focus instead on
the overdamped motions of classical
objects as they flow through two-dimensional periodic landscapes,
about which far less is known.
Such higher-dimensional periodic landscapes have shown exceptional
promise in a new category of sorting techniques.
Our discussion draws upon recent experimental realizations of this
process in which macromolecules or mesoscopic colloidal particles
are observed while moving through arrays of microfabricated posts \cite{duke97}
and through
regular arrays of optical traps \cite{korda02b,macdonald03,ladavac04}.
In both cases, particles' differing
interactions with the physical landscape
and their differing responses to the external driving force can cause
them to follow radically different paths, thereby providing a novel
basis for dispersing small fluid-borne objects into distinct fractions.

Section~\ref{sec:framework} introduces the theoretical framework for
describing driven objects' interactions with inhomogeneous environments in the
context of recent experimental realizations.
We then apply this in Sec.~\ref{sec:fringe} to the particularly simple case of
transport across a linear barrier or potential trench.  Such a landscape
can continuously sort mixtures of objects into two distinct fractions,
but with only algebraic sensitivity to properties such as size.
Generalizing to periodic landscapes in Secs.~\ref{sec:sine} and
\ref{sec:separable} leads
generally to fractionation with \emph{exponential} size selectivity.
Exploiting this exceptional resolution for practical separations may
be difficult, however, in the most straightforward implementations.
Other potential landscapes, such as a line of discrete potential wells,
discussed in Sec.~\ref{sec:linear}, 
offer exponential size selectivity with good prospects for practical
implementations.

\section{Motions Through Landscapes}
\label{sec:framework}

\subsection{The Equation of Motion}
Consider a Brownian particle moving, under the influence of a uniform driving force $\vec{F}_0$,
through the force field $\vec{F}(\vec{r})$ due to an inhomogeneous medium or landscape.
Its trajectory is described by the
Langevin equation \cite{risken89,mcquarrie00}
\begin{equation}
  \label{eq:smoluchowsky}
  \xi \, \frac{d \vecr}{dt} = \vec{F}(\vecr) + \vec{F}_0 + \vec{\Gamma}(t),
\end{equation}
where $\xi$ is the particle's viscous drag coefficient, and
$\vec{\Gamma}$ describes random thermal fluctuations.
This Langevin force satisfies $\langle \vec{\Gamma}(t) \rangle = 0$ and 
$\langle \vec{\Gamma}(t) \cdot \vec{\Gamma}(t + \tau) \rangle = \xi \, k_B T \, \delta(\tau)$
at temperature $T$, where $\delta(\tau)$ is the Dirac delta function.
A sphere of radius $a$ immersed in an unbounded fluid of viscosity $\eta$, for example, has
$\xi = 6 \pi \eta a$.

In the limit that $\vec{F}_0$ and $\vec{F}$ both greatly exceed the 
scale of thermal forces, $\vec{\Gamma}$,
the Langevin equation, Eq.~(\ref{eq:smoluchowsky}), reduces to a
first-order deterministic
equation of motion. 
This article focuses on two-dimensional systems, the simplest
case exhibiting novel behavior.
Even this deceptively simple system
yields surprising results, as we will see.

\subsection{The Driving Force}
In the particular case of 
fluid-borne colloidal particles,
a uniform driving force might be exerted by viscous drag,
by gravity,
or through electrophoresis, magnetophoresis, or thermophoresis.
Each of these plays a central role in practical
fractionation techniques \cite{wilson00}.
More generally, analogous results should be expected for such
related systems
as electrons flowing through a periodically-gated low-mobility
two-dimensional electron gas \cite{wiersig01},
magnetic flux quanta creeping through patterned type-II superconductors 
\cite{reichhardt00,reichhardt00c,reichhardt00d}
or Josephson junction arrays \cite{marconi01}, and atoms
migrating across crystal surfaces \cite{pierrelouis01}.

In some instances of practical interest, the driving force itself can be
modulated by the physical landscape, leading to additional interesting effects \cite{chou00}.
These, however, are beyond the scope of the present discussion.
Time-dependent driving forces also lead to exciting new phenomena,
but are not required for the effects we describe.
We consider the simplest case, where
the driving force $\vec{F}_0$ is both uniform and constant
and is oriented at
a fixed angle $\theta$ with respect to the 
landscape symmetry axis, here denoted $\hat x$.

In the absence of other influences, particles would travel
along the driving direction while dispersing diffusively
in the transverse direction.
Differential dispersion by transverse diffusion has proved useful
for continuously fractionating heterogeneous samples across
laminar flows in microfluidic channels
\cite{weigl99}.
Adding a modulated substrate opens up new modes of separating particles
according to their sizes, 
and can greatly improve the resolution
of such separations.

\subsection{Creating Landscapes}

Several approaches have been introduced in recent years 
for structuring potential energy landscapes
on molecular, macromolecular, and cellular levels.
Among these are arrays of lithographically defined
microscopic posts integrated into hermetically sealed fluidic channels, 
which provide a periodic and precisely tuned alternative to the gels used for
electrophoresis \cite{volkmuth92}.
Arrays of interdigitated electrodes \cite{rousselet94,gorretalini98} also have been
used to establish periodic potentials through dielectrophoresis.
The emphasis in these studies, however, has been on ratchet-like behavior
induced by time-dependent potentials.

More recently, techniques have been developed for tailoring extensive potential
energy landscapes using forces exerted by light.
The most capable of these exploit optical gradient forces,
meaning that dipole moments induced in illuminated objects respond to gradients
in the illumination's electric field.
Such forces are the basis for the single-beam optical trap known as an
optical tweezer \cite{ashkin86}, which acts as a potential energy
well for particles with appropriate optical properties.
More generally,
an extended optical intensity distribution will produce an 
associated potential energy landscape.

The most straightforward way to project periodic intensity profiles is to
create a standing-wave interference
pattern from two or more coherent beams of light.
Such patterns have come to be known as optical lattices,
particularly when applied to controlling the distributions and motions
of matter.
More general intensity patterns can be created with holographic
optical tweezers (HOT) \cite{dufresne98,dufresne01a,curtis02} or with the
generalized phase contrast (GPC) technique \cite{mogensen00},
which establish extended optical trapping patterns using computer-generated
holograms.

\subsection{Form Factors}

The physical landscape may be represented by a
function $I(\vecr)$ describing a potential-determining
property such as the local optical intensity.
An object's potential energy at $\vecr$ 
is determined, not only by $I(\vecr)$, but also by the
object's response to it.
For example, larger particles approaching a well-localized optical trap
encounter the trap's intensity gradients
at larger ranges than smaller particles.
The observation that
different objects passing through the same environment
experience different potential-energy
landscapes provides the foundation for the results that follow.

The effective potential may be expressed as the convolution
\begin{align}
  V(\vecr) & = (f \circ I)(\vecr) \label{eq:potential} \\
  & = \int f(\vec{x} - \vecr) \, I(\vec{x}) \, d^2x,
\end{align}
of the two-dimensional landscape, $I(\vecr)$, with a form factor
$f(\vecr)$ describing the object's interaction with the landscape.
In comparing to experimental realizations, we assume that contributions from
the form factor's third dimension have been integrated out.
If $I(\vecr)$ has a symmetry axis along the $\hat x$ direction, then
the associated force,
\begin{equation}
  \label{eq:landscape}
  \vec{F}(\vecr) = - \vec{\nabla} \, (f \circ I)(\vecr),
\end{equation}
generally does as well.
Convolving with $f(\vecr)$ broadens features in
$I(\vecr)$ by an amount that depends on the object's size,
shape, orientation, and composition.

In many cases of practical interest, the convolution in Eq.~(\ref{eq:potential})
is most easily performed using the Fourier convolution theorem:
\begin{equation}
\label{eq:convolution}
  (f \circ I)(\vecr) = {\cal F}^{-1} \{\tilde f(\vec{k}) \, \tilde I(\vec{k})\}
\end{equation}
where $\tilde f(\vec{k})$ and $\tilde I(\vec{k})$ are the Fourier transforms of
$f(\vecr)$ and $I(\vecr)$, respectively,
and $g(\vecr) = {\cal F}^{-1}\{\tilde g(\vec{k})\}$ 
denotes the inverse Fourier transform of
$\tilde g(\vec{k})$.
In some particularly simple cases, both $\tilde f(\vec{k})$ and $\tilde I(\vec{k})$
can be factored into components along the $\hat x$ and $\hat y$ directions, reducing
Eq.~(\ref{eq:convolution}) to a product of one-dimensional integrals.
In other cases, separable approximations for the form factor emerge as the leading-order
cumulant expansion of $\tilde f(\vec{k})$.

For example, the form factor for a uniform dielectric cube of side $a$ 
aligned with the $x$ axis and illuminated by collimated light is
\begin{equation}
  f(\vecr) = \alpha a \, \Theta \left( \frac{a}{2}-\abs{x} \right) 
  \Theta \left( \frac{a}{2}-\abs{y} \right)
\end{equation}
where $\Theta(x)$ is the Heaviside step function, and
\begin{equation}
  \alpha = 2 \pi \, \frac{\sqrt{\epsilon_0}}{c} \, 
  \left( \frac{\epsilon_0 - \epsilon}{\epsilon + 2 \epsilon_0}\right)
\end{equation}
describes the matter-light
interaction, in the quasi-static limit,
for a material of dielectric constant $\epsilon$ immersed
in a medium of dielectric constant $\epsilon_0$ \cite{bohren83}.
Both this geometry and the collimated light field are far simpler than would
be encountered in most real-world optical trapping implementations
\cite{maianeto00}, but serve to illustrate our approach.
A more complete treatment of optical forces also would 
incorporate polarization
effects, which cannot be captured in the present scalar theory.
Higher-order
effects such as Mie resonances \cite{bohren83} could be taken into account 
through $\alpha$,
but will be ignored in the current discussion.
Note that $\alpha$ is negative for a high-dielectric-constant material in a low-dielectric-constant medium;
such particles are drawn toward regions of high intensity.
Low-dielectric-constant particles, by contrast, are repelled by light.

The aligned cube's form factor is separable, 
with Fourier transform
\begin{equation}
  \tilde f(\vec{k}) = \alpha \, a^3 \, \tilde f_x(k_x a) \, \tilde f_y(k_y a).
\end{equation}
The individual components are readily shown to be
\begin{equation}
  \tilde f_x(ka) = \tilde f_y(ka) = \frac{\sin ka}{ka}.
\end{equation}
Their leading-order cumulant expansion,
\begin{equation}
  \tilde f_x(ka) = \tilde f_y(ka)
  \approx 
  \exp\left(- \frac{1}{6} \, k^2 a^2\right)
\end{equation}
for $ka < \pi$, demonstrates that the form factor's Fourier
transform depends sensitively on particle size for a given wavenumber.
Note that, as defined, $\tilde f_x(ka)$ and $\tilde f_y(ka)$ are dimensionless
and normalized to unity at $ka = 0$.

The form factor for a uniform dielectric sphere
of radius $a$ illuminated by collimated light of wavelength $\lambda > a$ is
\cite{bohren83},
\begin{equation}
  f(\vecr) = \alpha \, \sqrt{a^2 - r^2} \, \Theta(a-r),
\end{equation}
which is not separable.
The leading-order cumulant expansion of $\tilde f(\vec{k})$, however,
is separable, with
\begin{equation}
  \tilde f(\vec{k}) \approx \alpha \, \frac{2 \pi a^3}{3} \, \tilde f_x(k_x a) \,
  \tilde f_y(k_y a) \, ,
\end{equation}
where
\begin{equation}
  \tilde f_x(ka) = \tilde f_y(ka) =
  \exp\left(-\frac{1}{10} \, k^2a^2\right),
\end{equation}
for $k a < 8$.


More generally, an object's form factor is nonzero only over a limited domain, set
by its size.  
The corresponding Fourier transform thus depends strongly on $ka$,
within the appropriate range of wavenumbers.
We capture the ramifications of this boundedness
by adopting the separable Gaussian form
\begin{equation}
  \label{eq:gaussian}
  f(\vecr) = \alpha \, a \, \exp\left( - \frac{r^2}{2 a^2} \right),
\end{equation}
whose Fourier transform
\begin{equation}
  \tilde f (\vec{k}) = 2\pi \alpha a^3 \, \tilde f_x(k_x a) \, \tilde f_y(k_y a)
\end{equation}
 has components
\begin{equation}
  \label{eq:kgaussian}
  \tilde f_x(ka) = \tilde f_y(ka) = \exp\left( - \frac{1}{2} k^2 a^2 \right).
\end{equation}
Which wavenumbers come into play depends on the landscape, $I(\vecr)$.
The following Sections explore a few particularly
effective choices.

\section{Linear Fringes}
\label{sec:fringe}

In part to motivate a discussion of periodic potential energy
landscapes, we first consider how objects traverse a single
trench or barrier arranged at an angle to the driving force.
This kind of landscape
may be realized, for example, by creating a linear optical trap 
with a cylindrical
lens or a diffractive line generator.
Because such an optical landscape can act as either a barrier or a trench,
depending on the sign of $\alpha$, we will refer to both as fringes.
In either case, a fringe aligned with the $\hat x$ axis
inhibits
transport in the transverse direction.
We model the
landscape as
a Gaussian profile of intrinsic width $w$,
\begin{equation}
  \label{eq:intens1trench}
  I(\vecr) = I_0 \, \exp\left(-\frac{y^2}{2 w^2}\right).
\end{equation}
Using the object's form factor as defined in Eq.~(\ref{eq:gaussian}),
the associated potential is
\begin{equation}
  \label{eq:pot1trench}
  V(\vecr) = 2\pi\alpha I_0 \,
  \frac{a^3 w}{\sigma(a)} \, \exp\left(-\frac{y^2}{2 \sigma^2(a)}\right).
\end{equation}
The fringe's apparent width to a particle of size $a$
is broadened to $\sigma(a) = \sqrt{a^2 + w^2}$.

In the limit that thermal forces may be ignored,
the equations of motion reduce to the deterministic form
\begin{align}
  \label{eq:eom1trenchx}
  \frac{dx}{dt} & = v_0 \, \cos \theta \\
  \label{eq:eom1trenchy}
  \frac{dy}{dt} & = \xi^{-1} \, F_y(y) + v_0 \, \sin \theta,
\end{align}
where the landscape-free drift speed is $v_0 = \xi^{-1} F_0$ and
\begin{equation}
  \label{eq:trenchFy}
  F_y(y) = 2\pi \, \alpha I_0 \, \frac{a^3 w}{\sigma^3(a)} \,
  y \, \exp\left(- \frac{y^2}{2 \sigma^2(a)}\right).
\end{equation}

The landscape's restoring force $F_y(y)$ reaches a maximum at a distance $y = \ymax$
from the fringe's axis, with $\ymax = \sigma(a)$ for our particular example.
If $F_0 \sin \theta > F_y(\ymax)$ then a particle can cross the
barrier.
Such particles may be said to \emph{escape} the barrier.
By contrast, particles for whom the barrier is insurmountable
travel unimpeded along the $\hat x$ direction
at speed $v_x = v_0 \cos \theta$.
Such particles are said to be \emph{locked in} to the landscape.

The marginal angle $\theta_m$ at which an object just barely
remains locked in to the barrier
determines which objects are deflected and which are not.
The dependence of $\theta_m$ on particle size and other
characteristics establishes 
the sensitivity of the
sorting technique.
Referring to Eqs.~(\ref{eq:eom1trenchy}) and (\ref{eq:trenchFy}),
the condition for locked-in transport,
\begin{align}
  \label{eq:condition1}
  \sin \theta & \leq \sin \theta_m \equiv \frac{F_y(\ymax)}{F_0} \\
  & = \frac{\abs{\alpha} \, I_0}{F_0} \, \frac{2 \pi}{\sqrt{e}} \, 
  \frac{a^3 w}{a^2 + w^2},
  \label{eq:condition1a}
\end{align}
applies both to
attractive trenches ($\ymax=+\sigma$) and repulsive barriers ($\ymax=-\sigma$).
The general result, Eq.~(\ref{eq:condition1}), applies even if $f(\vecr)$ is
not separable because, in this case at least, $I(\vecr)$ is independent of $x$.

The particular result in Eq.~(\ref{eq:condition1a}) shows that
the marginal lock-in angle
depends only algebraically on size, and only
linearly on other properties through $\alpha$.
This is neither better nor worse that the performance offered
by other established
techniques such as gel electrophoresis or flow-field fractionation \cite{wilson00}.
One substantial benefit offered by selective transport across a fringe is
its ability to
process a continuous stream of
objects rather than being restricted to discrete batches.
The selected fraction, moreover, can be tuned continuously, for example
by adjusting $I_0$, $F_0$, $w$, or $\theta$.
Optical implementations also can be optimized by varying the
wavelength of light, in which case resonances might be exploited
as a complementary mechanism for size separation.

Although a single fringe's performance is somewhat lackluster,
one might expect multiple fringes to fare better.
The first step along this direction is to consider
a pair of parallel Gaussian fringes.
The effective potential is the sum of
two single-fringe potentials:
\begin{multline}
  \label{eq:pot2trench}
  V(\vecr) = 2\pi\alpha I_0 \,
  \frac{a^3 w}{\sigma(a)} \times \\
  \left[ \exp\left(-\frac{\left(y+b/2\right)^2}{2 \sigma^2}\right)  
+ \exp\left(-\frac{\left(y-b/2\right)^2}{2 \sigma^2}\right) \right],
\end{multline}
where $b$ is the fringe separation.  
If $b < \sigma$, the two fringes overlap enough that the
landscape resembles a single, broadened fringe.
Again, no more than algebraic selectivity should be expected.
In the opposite limit, $b \gg \sigma$, the fringes are independent, and 
particles cross the double barrier with the same facility with which they
cross one.
Neither of these cases offers benefits over the single fringe.

For intermediate $b$, on the other hand, the landscape consists of 
two unequal barriers, the smaller of which lies between the two fringes.
The smaller barrier's height depends strongly on $b/\sigma$, which, in turn,
depends on the particle size $a$.
This lower intermediate barrier does not affect the fringes' overall
ability to separate objects, which is dominated by the larger barrier.
It suggests though possibility, that
transport across
$N$ overlapping fringes
could be highly sensitive to particle size.
Those particles not able to
jump the inter-fringe barriers will be locked in and swept aside
while others
will hop from one fringe to the next across the field.
Highly selective sorting is thus possible if edge effects due to the 
first or last
fringe (in the case of trenches or barriers, respectively) can be circumvented.

\section{Sinusoidal Landscapes}
\label{sec:sine}

\begin{figure}[htbp]
  \centering
  \includegraphics[width=0.9\columnwidth]{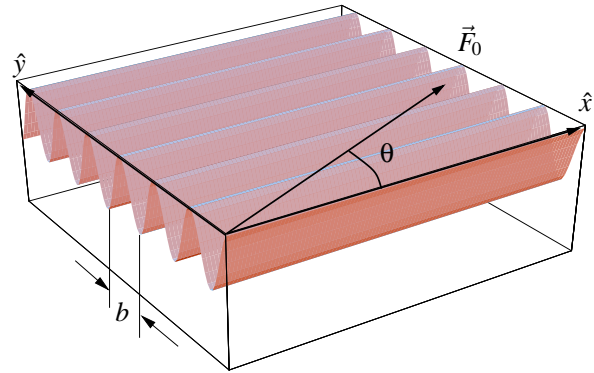}
  \caption{Schematic representation of a sinusoidal landscape, modulated
  along the $\hat y$ direction.}
  \label{fig:sinusoid}
\end{figure}
The foregoing discussion suggests that a periodically modulated landscape
inclined at an angle to the driving force might be more effective
than a single fringe at sorting objects by size.
To make this more concrete, and to illuminate the role of periodicity, we consider the
simplest and most instructive example of such a landscape,
a sinusoid in the $\hat y$ direction:
\begin{equation}
  \label{eq:sine}
  I(\vecr) = I_0 \, \cos (k_0 y),
\end{equation}
shown schematically in Fig.~\ref{fig:sinusoid}.
Apart from its mathematical
simplicity, this landscape has the advantage of being readily implemented experimentally.
In the case of optical forces, a sinusoidal pattern can be created
by interfering two coherent laser beams, with the spatial wavenumber $k_0$
determined by the optical wavelength and
the angle between the beams.  Such an interference pattern is
known as a one-dimensional optical lattice \cite{chiou97}, and is commonly used to
control and distribute
cold atoms.
More recently, optical lattices have
been used to separate populations of colloidal particles on the basis
of their sizes and indices of refraction \cite{macdonald03}.

The key to such a potential's efficacy is in its Fourier transform:
\begin{equation}
  \tilde I(\vec{k}) = (2 \pi)^2 \, I_0 \, \delta(k_x) \, \delta (k_y - k_0).
\end{equation}
Convolution according to equation \ref{eq:convolution} then results in the potential landscape
picking out the component of the form factor's Fourier transform at wavenumber $k_0$:
\begin{align}
  V(\vec{r}) & = I_0 \, \int \tilde f(\vec{k}) \, 
  \delta(k_x) \, \delta (k_y - k_0) \, \cos(k_x x) \cos(k_y y) \, d^2k\\
  & = I_0 \, \tilde f(0,k_0) \, \cos(k_0 y).
\end{align}
Assuming a separable form for $\tilde f(\vec{k})$ such
as Eq.~(\ref{eq:kgaussian}),
\begin{align}
  V(\vecr) & = I_0 \, \tilde f_y(k_0 a) \, \cos (k_0 y) \label{eq:sinegeneral} \\
  & = \alpha \sqrt{2\pi} I_0 \, a^3 \, 
  \exp\left(- \frac{1}{2} \, k_0^2 a^2\right) \, \cos (k_0 y).
  \label{eq:sine-potential}
\end{align}

For particle sizes such that $k_0 a > \sqrt{3}$,
the amplitude of the landscape's sinusoidal
modulations now depends strongly on particle size
through the wavenumber dependence of the form factor.
The particular form in Eq.~(\ref{eq:sine-potential}) reflects
our choice of a Gaussian form factor in Eq.~(\ref{eq:gaussian}).
However, the arguments leading to this choice reveal that comparable
results should be obtained quite generally for particles whose size
is smaller than the wavelength of the physical
landscape's undulations ({\it e.g.}, for spherical particles, for
$k_0 a < 8$).

\subsection{Deterministic Limit}

A particle's trajectory through the sinusoidal landscape
is described in the deterministic limit by
Eqs.~(\ref{eq:eom1trenchx})
and (\ref{eq:eom1trenchy}), with
\begin{equation}
  F_y(y) = \alpha \sqrt{2\pi} I_0 k_0 a^3 \, 
  \exp\left(- \frac{1}{2} \, k_0^2 a^2\right) \, \sin (k_0 y).
\end{equation}
Non-separable form factors again yield similar results,
because $I(\vecr)$ is independent of $x$.
Because $\sin(k_0 y) \le 1$,
particles become locked into the
$\hat x$ direction for orientations satisfying
\begin{equation}
  \label{eq:condition2}
  \sin \theta \leq \sin \theta_m = \frac{\abs{\alpha} \, I_0}{F_0} \,
  \sqrt{2 \pi} k_0 a^3 \, \exp\left(- \frac{1}{2} \, k_0^2 a^2\right).
\end{equation}
By contrast to Eq.~(\ref{eq:condition1a}),
this reflects an exceptional sorting sensitivity:  whether or not 
a particle becomes entrained by the fringes
depends \emph{exponentially} on particle
size for $k_0 a > 1$.
Among established fractionation schemes, only affinity chromatography
offers comparable selectivity \cite{wilson00}, and this can
operate only on discrete samples of a limited class of macromolecules.
Fractionation in a sinusoidal landscape,
by contrast, can operate on continuous
sample streams and can be implemented for a wide range of sample types.

Equations~(\ref{eq:eom1trenchx}), (\ref{eq:eom1trenchy}), and (\ref{eq:sine-potential})
can be directly integrated for this
simple landscape.  The motion in the $\hat x$ direction is trivial:
\begin{equation}
  \label{1dfringe-x}
  x(t) = v_0 \cos \theta \, t .
\end{equation}
If the particles are locked in (\emph{i.e.}, if Eq.~(\ref{eq:condition2}) 
is satisfied),
then the particles make no progress in the $\hat y$ direction, 
and $y(t)$ is constant in steady state.  Otherwise,
integration gives
\begin{equation}
  y(t) = \frac{2}{k_0} \,
  \arctan\left[ \sqrt{\frac{\sin\theta + \eta}{\sin\theta - \eta}} \,
    \tan \left( \frac{k_0 v_0 t}{2} \, \sqrt{\sin^2\theta - \eta^2}\right) \right],
\end{equation}
where the relative strength of the landscape's modulation is
\begin{equation}
  \label{eq:etasin}
  \eta(a) = \frac{I_0 \, k_0 \tilde f(0, k_0a)}{F_0}.
\end{equation}
This motion can be seen by inspection to be periodic, 
with the period
\begin{equation}
  T = \frac{2 \pi}{k_0 v_0 \sqrt{\sin^2 \theta - \eta^2}}
\end{equation}
corresponding to the time to advance
one fringe spacing,
$b = 2 \pi/ k_0$.
The mean velocity in the $\hat y$ direction is thus given by
\begin{equation}
  \label{1dfringe-y}
  \langle v_y \rangle = v_0 \sqrt{\sin^2\theta - \eta^2}.
\end{equation}
On average, the particle travels at an angle $\psi$
to the $\hat x$ axis, given by
\begin{equation}
  \label{1Ddeterministic}
  \tan \psi = \frac{\langle v_y\rangle}{\langle v_x \rangle} =
  \begin{cases}
    0 , & \sin \theta < \eta \\
    \frac{\sqrt{\sin^2\theta-\eta^2}}{\cos\theta} , & \sin \theta > \eta
    \end{cases} \, .
\end{equation}

\begin{figure}[htbp]
  \centering
  \includegraphics[width=\columnwidth]{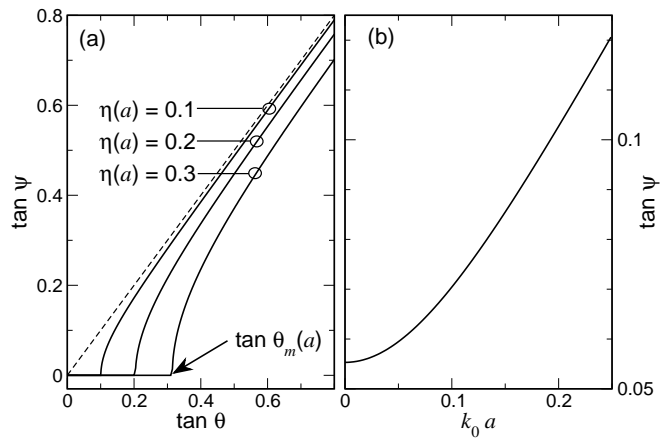}
  \caption{(a) Travel direction as a function of orientation for an inclined
  sinusoidal landscape as a function of orientation for fixed size and
  $\eta(a) = 0.1$, 0.2, and 0.3.
  Trajectories are locked in to $\psi(\theta) = 0$ for $\theta \le \theta_m$.
  The diagonal dashed line indicates the result with no landscape.
  (b) Deflection angle as a function of particle size $a$ at 
  fixed driving orientation $\tan \theta = 0.441$, assuming
  $\eta_0 = 0.4$, independent of $a$.}
  \label{fig:detsin}
\end{figure}
The deflection angle is plotted in Fig.~\ref{fig:detsin}(a) 
as a function of the of the driving force's orientation $\theta$
for various values of the normalized potential $\eta(a)$.
The direction in which particles flow increases from
$\psi = 0$ as the driving force's orientation crosses the condition
for marginal lock-in,
$\theta_m = \arcsin \eta$.
At steeper angles, $\psi$ approaches $\theta$.

While this result is quite general,
we can make the dependence on particle size more explicit by assuming
the following functional form for $\eta(a)$, implied 
by Eqs.~(\ref{eq:gaussian}) and (\ref{eq:etasin}):
\begin{equation}
\eta(a) = \eta_0 \exp \left( -\frac{1}{2} k_0^2 a^2 \right) ,
\end{equation}
where $\eta_0 = 2 \pi \alpha a^3 I_0 k_0 / F_0$.  Fig.~\ref{fig:detsin}(b)
shows the resulting dependence of deflection angle on particle size, if we
assume that the driving and trapping forces are adjusted such that $\eta_0$
is a constant, independent of $a$.
It can be seen that
particles which are not locked in to the fringes at $\psi(a) = 0$ are
fanned out into
various directions, depending on their size.  

Unlike the case of the single fringe, where a particle either 
flows along the fringe or else travels in the driving direction,
the sinusoidal landscape's continuous dispersion distributes heterogeneous
samples into multiple fractions,
but also limits the achievable
size resolution.
The fraction dispersed into a finite angular range $\Delta \psi$ around $\psi$
includes an associated range of sizes
\begin{equation}
  \label{eq:sineresolution}
  \Delta a \approx
  \left(\frac{\partial \psi}{\partial a} \right)^{-1} \, \Delta \psi,
\end{equation}
which, for the locked-in fraction at $\psi = \Delta \psi / 2 \ll 1$ is
\begin{equation}
  \label{eq:sineresolution2}
  \Delta a \approx \cos^2 \theta \,  \left(\frac{\partial \eta^2}{\partial a} \right)^{-1} \,
  \Delta \psi^2.
\end{equation}
Thus, the exponential size selectivity implied by Eq.~(\ref{eq:condition2})
can be lost in the exponentially wide collection window imposed by $\eta(a)$
on practical implementations.
This performance cannot be improved by passing
the set of particles through the fringes a second time, 
because of the fixed relationship between
$\Delta a$ and $\Delta \psi$.

Although single-stage fractionation by a sinusoidal landscape
yields broad size distributions,
a narrow range of particle sizes still can be 
captured by using the following, two-step
process.
The deflection angle is first set such that all particles larger than a certain
size $a_2$ will be locked in.  These locked-in particles are discarded, and the remaining
particles are sent through a second potential landscape, with a different deflection angle,
chosen such that all particles larger than a second size $a_1 < a_2$ are locked in.
Only the locked-in particles from this second stage are then retained, so that all of the remaining
particles have sizes in the range $\left[ a_1 , a_2 \right]$, which can be made as
small as desired.

With periodicity providing the essential ingredient for achieving exponential
size selectivity, it might be expected that any periodic landscape would do.
Unfortunately, this is not necessarily so.
We already have demonstrated in Sec.~\ref{sec:fringe} 
that an array of well-separated Gaussian fringes offers only algebraic, rather than exponential,
size selectivity.
A more general periodic landscape with wavelength $2 \pi / k_0$ can be expanded
as a Fourier series:
\begin{equation}
  I(\vecr) = I_0 \,\sum_{n=0}^\infty \beta_n \sin (n k_0 y),
\end{equation}
with Fourier coefficients $\beta_n$.  
If one of these coefficients is significantly larger than all the others,
then the equations of motion can be approximated by
Eqs.~(\ref{eq:eom1trenchx}) and (\ref{eq:eom1trenchy}),
and equivalently good
size selectivity will be obtained.
If, on the other hand, no single component dominates,
then the superposition will not necessarily perform so well.

\subsection{Biased Diffusion}

Modelling thermal effects is reasonably straightforward for the
sinusoidal landscape.
In this case, the Langevin
equation (Eq.~(\ref{eq:smoluchowsky})) is most readily solved by transforming it into a
Fokker-Planck equation for the probability density $\rho(\vecr,t)$ of finding particles
at position $\vecr$ at time $t$.  If inertial effects are negligible, the Fokker-Planck 
equation for motion transverse to the fringes reduces
to a Smoluchowski equation 
\cite{risken89},
\begin{equation}
  \label{fock-her-planck}
  \partial_t \rho(y,t) + \partial_y S(y,t) = 0,
\end{equation}
where the probability current is
\begin{equation}
  S(y,t) = \xi^{-1} \left( F(r) - k_B T \partial_y \right) \rho (y,t) \, .
\end{equation}
In this equation, $F(r)$ is the total force on the particle, including the driving force and
the force due to the potential landscape, $T$ is the temperature, and $k_B$ is Boltzmann's
constant.

\begin{figure}[htbp]
  \centering
  \includegraphics[width=\columnwidth]{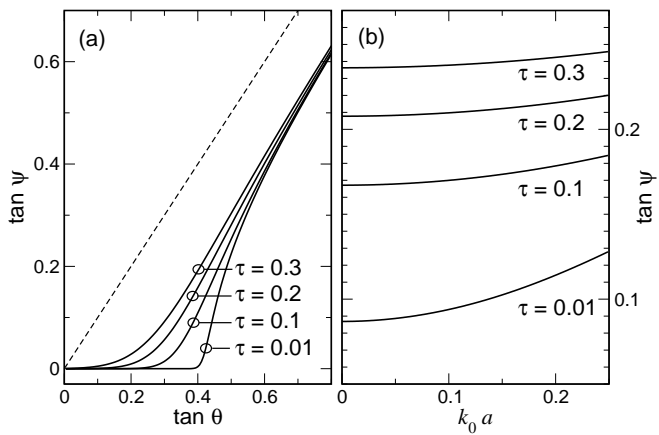}
  \caption{(a) Deflection as a function of orientation at finite temperatures,
    $\tau = 0.01$, 0.1, 0.2 and 0.3, assuming $\eta = 0.4$.
    The diagonal dashed line indicates the result with no landscape.
    (b) Dependence of the travel direction on particle size $a$ for $\eta_0 = 0.4$,
    $\tan \theta = 0.441$ and the same set of temperatures.  Raising the temperature
    weakens the size dependence of $\psi(a)$, and thus reduces the selectivity.}
  \label{fig:1d}
\end{figure}
Following Ref.~\cite{risken89}, Eq.~(\ref{fock-her-planck}) can be solved in the steady-state
limit by taking $S(y,t) = S(y)$, independent of $t$.
The resulting average drift 
velocity $\langle v_y \rangle$ for the sinusoidal potential of 
Eq.~(\ref{eq:sine-potential}) is given by
\begin{equation}
  \label{drift}
  \frac{\langle v_y \rangle}{v_0 \sin\theta} = 1 + \frac{2 \sin\theta}{\eta} \, 
  \Im \left[S_1\left(\tau,\frac{\sin \theta}{\eta}\right)\right]
\end{equation}
where, as before, $\eta = k_0 I_0 \tilde f(0,k_0 a) / F_0$, and we have introduced the
normalized temperature $\tau = k_B T/V_0$.  
The function $S_1(\tau,x)$ is defined recursively in terms of a
continued-fraction expansion,
\begin{equation}
  S_n\left(\tau,x\right) = 
  \frac{1/4}{\tau + i n \, x + 
    S_{n+1}\left(\tau, x\right)},
\end{equation}
which converges rapidly with increasing order $n$.

The average velocity in the $\hat x$ direction is unchanged 
from the zero-temperature case.  The mean
deflection angle $\psi$ is thus given by
\begin{equation}
  \label{deflect1D-thermal}
  \tan \psi = \tan \theta \left( 1 + \frac{2 \sin\theta}{\eta} \, 
    \Im \left[S_1\left(\tau, \frac{\sin \theta}{\eta}\right)\right] \right) \, .
\end{equation}
This is plotted in Fig.~\ref{fig:1d}(a) as a function of the angle $\theta$ of the driving force, 
for a fixed value
of the normalized potential $\eta(a)$, and for various 
values of the normalized temperature $\tau$.
It can be seen
that the effect of increasing temperature is to smooth out the transition 
between the locked-in and freely flowing
states of motion.  In the zero-temperature limit, the deflection angle is zero for all angles $\theta < \arcsin\eta$.
For finite temperatures, the mean deflection angle is non-zero even in this ``locked-in" state:
the particles have a finite probability per unit time of 
being driven over the inter-fringe barrier by thermal fluctuations,
and thereby advancing in the $\hat y$ direction.


The benefits of operating in the deterministic regime, in which thermal forces are
negligible, can be seen in Fig.~\ref{fig:1d}(b), where the deflection angle is shown as a
function of the particle size $a$, for a fixed orientation $\theta$,
over a range of temperatures.
Contrary to previous assertions that thermal effects can enhance
size selectivity \cite{macdonald03}, the only effect of thermally-assisted
hopping in this system is to diminish the sorting resolution.

In other words, achieving high-sensitivity sorting in practice
will require that the thermal energy scale be small compared to the
landscape's modulation.
In a real-world implementation, 
increasing the depth of modulation $I_0$ of the physical
landscape 
will often be more practical than
decreasing the temperature.
Retaining the same lock-in conditions then requires that the
driving force $F_0$ to be increased proportionately.
The practical limit on the achievable sorting efficiency will then be set by the
maximum driving force or depth of modulation that can be obtained.  
For example, the limitation for an optically-implemented landscape
is the available laser power.

\section{Separable Two-Dimensional Landscapes}
\label{sec:separable}

\begin{figure}[htbp]
  \centering
  \includegraphics[width=0.9\columnwidth]{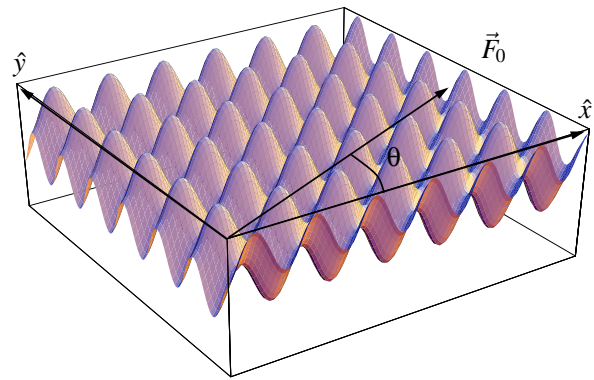}
  \caption{Schematic representation of a landscape sinusoidally modulated
  in both the $\hat x$ and $\hat y$ directions according to 
  Eq.~(\ref{eq:sine2D}).}
  \label{fig:fringe2D}
\end{figure}
The one-dimensional potential of Eq.~(\ref{eq:sine}) is one of the few landscapes that allows for exact
solutions of the equations of motion.  Two-dimensional landscapes can be solved analytically
only if the potential can be written as a sum of modulations in the $\hat x$ and $\hat y$ directions.
In particular,
we can consider separate sinusoidal modulation with the same period in the two directions:
\begin{equation}
  \label{eq:sine2D}
  I(\vecr) = I_0 \, \left[ \sin (k_0 y) + \sin (k_0 x) \right],
\end{equation}
shown schematically in Fig.~\ref{fig:fringe2D}.
This landscape is interesting mainly because it 
leads to decoupled equations of motion:
\begin{align}
\label{eq:2Dfringe}
  \frac{dx}{dt} & = \xi^{-1} V_0 k_0 \, \cos (k_0 x) + v_0 \, \cos \theta + \gamma(t) \\
  \frac{dy}{dt} & = \xi^{-1} V_0 k_0 \, \cos (k_0 y) + v_0 \, \sin \theta + \gamma(t),
\end{align}
where $\gamma(t) = \Gamma/\xi$.
Nevertheless, such a landscape could be implemented
experimentally using optical forces.  
For example, mutually incoherent pairs of laser beams intersecting
at right angles
would lead to a potential of the form given in Eq.~(\ref{eq:sine2D}),
as would pairs with orthogonal polarization.

Since the motions in the $\hat x$ and $\hat y$ directions are independent in this case, 
the same exponential sensitivity to particle size, Eq.~(\ref{eq:condition2}) is 
obtained, in the absence of thermal forces.
As well, the same integration can be used to determine the average deflection angle for 
free-flowing particles, analogous to Eq.~(\ref{1Ddeterministic}):
\begin{equation}
  \label{2Ddeterministic}
  \tan \psi = \frac{\sqrt{\sin^2\theta - \eta^2}}{\sqrt{\cos^2\theta - \eta^2}}.
\end{equation}
Similarly, for finite temperatures, the mean deflection angle is given by
\begin{equation}
  \label{deflect2D-thermal}
  \tan \psi = \tan \theta \;
  \frac{1 + \frac{2 \sin\theta}{\eta} \, 
    \Im \left[S_1\left(\tau, \frac{\sin \theta}{\eta}\right)\right]}
  {1 + \frac{2 \cos\theta}{\eta} \, 
    \Im \left[S_1\left(\tau, \frac{\cos \theta}{\eta}\right)\right]}.
\end{equation}

\begin{figure}[htbp]
  \centering
  \includegraphics[width=\columnwidth]{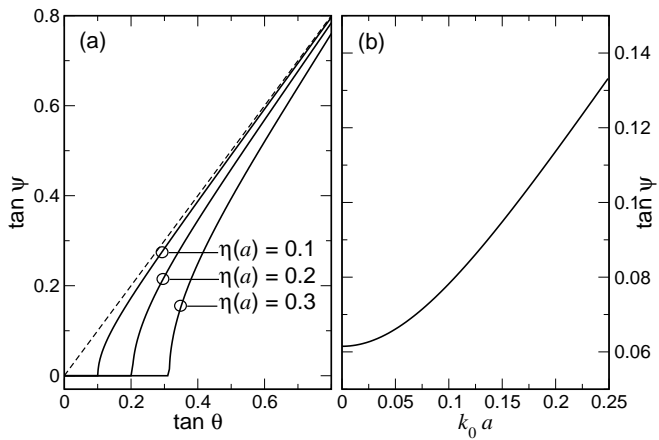}
  \caption{(a) Deflection as a function of orientation for a separable
  two-dimensionally modulated landscape at $\eta(a) = 0.1$, 0.2, and 0.3.
  The diagonal dashed line indicates the result with no landscape.
  (b) Size dependence of the deflection angle for $\eta_0 = 0.4$ and
  $\tan \theta = 0.441$.}
  \label{fig:2D}
\end{figure}
The zero temperature
deflection angle is plotted in Fig.~\ref{fig:2D} as a function of the
angle $\theta$ of the driving force, for a fixed value of the
normalized potential $\eta$.
Also plotted is $\psi$ as a function of particle
size $a$ for a fixed $\theta$.  The results can be seen to be similar
to those obtained with one-dimensional fringes.
In other
words, no qualitative difference is obtained in this case by
modulating in two directions rather than just one.

In order to see new effects of increased dimensionality, it is necessary to consider
landscapes that cannot be separated into one-dimensional terms;
{\em i.e.}, landscapes where the motion in one
dimension depends on the position in the other.  Analytical
solutions are not available for such landscapes.  However, it is
possible to develop limiting arguments that
illustrate the novel features of transport in such landscapes, including
the continued possibility for sorting that is exponentially sensitive
to particle size.

\section{Linear Trap Arrays}
\label{sec:linear}

\subsection{Periodically modulated fringe}
Combining aspects of Secs.~\ref{sec:fringe} and \ref{sec:sine}, 
we next consider landscapes that are 
uniform outside a bounded region in the $\hat y$ direction, and are
periodically modulated in the $\hat x$ direction.
This features the clean separations of the continuous barrier with the
exponential selectivity of sinusoidal landscapes.
It also provides a straightforward example of the surprising complexity of
non-separable landscapes.

The simplest exemplar is the modulated line,
\begin{equation}
  \label{eq:sineline}
  I(\vecr) = I_0 \, A(y) \, \frac{1 + s \cos(k_0 x)}{1 + s},
\end{equation}
where $A(y)$ describes the transverse profile and is peaked at
$A(0) = 1$.  Here, the factor $s$
controls the depth of modulation
along the line and falls in the range $0< s < 1$.
Such an array can be realized, for example, with 
a linear array of discrete optical tweezers.
Choosing $s > 1$ would correspond to alternating potential wells
and barriers,
which also could be implemented optically, for instance with two different
wavelengths of light.

At a given driving orientation $\theta$, 
objects are either locked in to the array and deflected, or else escape
into the driving direction.  This is unlike the sinusoidal landscape, for which
even the particles that are not locked in are deflected away from
the driving direction.  
Collection of the desired fraction should thus be more straightforward for
the linear array of traps.

The equations of
motion for objects driven deterministically through $I(\vecr)$
at angle $\theta$ are
\begin{align}
  \label{eq:eom2dx}
  \frac{dx}{dt} & = \xi^{-1} \, F_x(\vecr) + v_0 \, \cos \theta \\
  \label{eq:eom2dy}
  \frac{dy}{dt} & = \xi^{-1} \, F_y(\vecr) + v_0 \, \sin \theta,
\end{align}
where the components of the substrate-mediated force reflect a convolution
with a particle's form factor.
We will assume
the particles' form factors to be separable, as in Eq.~(\ref{eq:gaussian}),
thereby sacrificing some generality in favor of clarity, so that
\begin{align}
  F_x(\vecr) & = 
  \alpha I_0 k_0 a^2 \, \bar A(y) \, 
  \frac{s \tilde{f}_x(k_0 a) \sin(k_0 x)}{1 + s} \\
  F_y(\vecr) & =
  - \alpha I_0 a^2 \, 
  \partial_y \bar A(y) \, \frac{1 + s \tilde{f}_x(k_0 a) \cos(k_0 x)}{1 + s},
\end{align}
where the effective transverse profile is $\bar A(y) = (f_y \circ A)(y)$.
Even this simplified set of coupled equations is highly nonlinear and cannot be 
integrated directly.
Instead, we resort to limiting arguments to determine 
when particles become locked in and when they escape.
These estimates provide the basis for our
claim that sorting by inclined arrays of traps or barriers
can offer exponential size selectivity.

As for the uniform fringe in Sec.~\ref{sec:fringe}, the restoring force
$F_y(\vecr)$ attains its maximum value
along $y = \ymax$, with $\ymax > 0$ for attractive wells and
$\ymax < 0$ for repulsive barriers.
For this separable model, \ymax is a solution to $\partial^2_y \bar A(y) = 0$.
In more general non-separable systems, it varies with position $x$ along the array.
In either case, this threshold depends on the object's geometry and composition
through the form factor $f(\vecr)$.

A particle's trajectory must cross $y = \ymax$ if it is to escape the line of traps.
This requires there to be at least some points $x$ along the array where
$F_0 \sin \theta > F_y(x,\ymax)$.
Limits on this condition can easily be established, with
\begin{equation}
  \label{eq:thetamax}
  F_0 \sin \theta \ge \max_x\{F_y(x,\ymax)\}
\end{equation}
ensuring that every trajectory escapes and
\begin{equation}
  \label{eq:thetamin}
  F_0 \sin \theta \ge \min_x\{F_y(x,\ymax)\}
\end{equation}
opening up the possibility that at least some trajectories might.
The associated bounds on $\theta_m$, the marginally locked-in angle, are
\begin{equation}
  \label{eq:bounds}
  \eta(a) \, \frac{1-s \tilde{f}_x(k_0 a) }{1+s} < \sin\theta_m < 
  \eta(a)\, \frac{1+s \tilde{f}_x(k_0 a) }{1+s},
\end{equation}
where $\eta(a) = \alpha k_0 a^2 \bar A(\ymax) / F_0$.
With this definition, $\eta(a)$ depends
only weakly on $a$, as in Sec.~\ref{sec:fringe}.

The limit of weak modulation, $s = 0$, once again yields
Eq.~(\ref{eq:condition1a}), the result for a continuous barrier or trench.
Similarly, since $\lim_{a \rightarrow \infty} \tilde f_x (k_0 a) = 0$,
large particles with $k_0 a > 1$ are not significantly affected by the 
modulation.
Smaller particles encountering a deeply modulated line, $s\rightarrow 1$,
are more interesting.  Unfortunately, the simple bounds in Eq.~(\ref{eq:bounds})
have no predictive power in this range, because 
$\lim_{a \rightarrow 0} \tilde{f}_x(k_0 a) = 1$ and Eq.~(\ref{eq:bounds}) reduces
to $0 < \sin \theta_m < \eta(a)$.

This is not to say that exponential selectivity is lost in this range, but rather
that a more detailed analysis is required to ascertain when it can be attained.
To illustrate the possibility of achieving exponential sensitivity,
we consider an experimentally realizable landscape consisting
of a line of discrete optical tweezers \cite{ladavac04}, which we model as a
line of discrete Gaussian wells.

\subsection{Line of Gaussian Wells}
After convolution with a Gaussian form factor, a single well of intrinsic
width $w$ takes the form
\begin{equation}
  V_1(\vec r) = - V_0 \, \exp \left( - \frac{r^2}{2 \sigma^2} \right)
\end{equation}
with $\sigma^2(a) = w^2 + a^2$.
This should not be mistaken for an accurate model of an optical tweezer's
potential well, but rather as a tractable model whose behavior
approximates that observed in actual optical traps.
A line of such wells separated by a distance $b$ results in the
potential energy landscape
\begin{equation}
  \label{eq:lgpotential}
  V(\vec r) = V_0(a) \sum_{n} \exp\left( - \frac{(\vec r - n b \hat x)^2}{2 \sigma^2(a)} \right).
\end{equation}

Any trajectory locked in to this periodic landscape will itself be
periodic in $x$.
This means that such a trajectory passes through a sequence of turning points at which
$\partial_x y(x,t) = \partial_t y(x,t) = 0$.
Any trajectory lacking such turning points cannot be locked in, and so must
escape from the line of potential wells.
Turning points come in two varieties: those where
particles make their nearest approaches to the wells' centers, and those corresponding
to their furthest excursions from the line of traps.
Particles can escape when the latter type disappear.

For small to moderate driving angles $\theta$, the more distant turning 
points occur near the midplanes between traps, where the restoring force
is weakest.
Considering the influence of just two traps (appropriate for $b > \sigma$),
centered at $x = 0$ and $x = b$,
this suggests the point of escape will be near $x = b/2$ and
$y = \sigma$.
Expanding around this point yields
\begin{equation}
  \label{eq:lgcriterion}
  \sin \theta_m \lesssim \eta(a) \, \exp\left( - \frac{b^2}{8 \sigma^2} \right).
\end{equation}
where $\eta(a) = (2/\sqrt{e}) \, V_0/( \sigma F_0)$ 
measures the traps' strength relative to the driving force.
Because Eq.~(\ref{eq:lgcriterion}) is an upper bound, 
no locked-in trajectories can occur for $\theta > \theta_m$.
Equation~(\ref{eq:lgcriterion}) therefore establishes the exponential
size dependence of particles' deflection.

\begin{figure}[htbp]
  \centering
  \includegraphics[width=0.9\columnwidth]{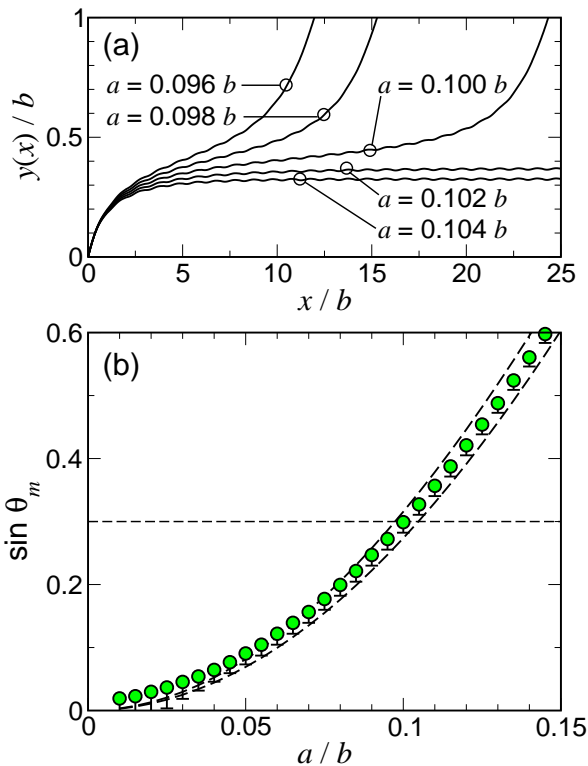}
  \caption{(a) Trajectories calculated according to Eqs.~(\ref{eq:eom2dx})
    and (\ref{eq:eom2dy}) for the line
    of Gaussian wells described by Eq.~(\ref{eq:lgpotential}).  The
    wells are separated by distance $b$ and have intrinsic width $w = 0.4\, b$.
    Their effective width is $\sigma = \sqrt{w^2 + a^2}$, where $a$ is the
    radius of a sphere flowing through the array.  The effective potential
    well depth is $\eta(a) = (2/\sqrt{e}) \, V_0/(\sigma F_0) = 9.7 \,\sigma^2/a^2$.
    With the driving force oriented at $\theta = 17.5^\circ$, spheres with radii larger
    than $a = 0.1~b$ are locked into the line.
    (b) Dependence of the marginally locked-in deflection angle $\theta_m$ on 
    radius, $a$.
    The lower dashed curve is the prediction of Eq.~(\ref{eq:lgcriterion}) and the upper
    from Eq.~(\ref{eq:bounds}).  The dashed line indicates the orientation along which
    the data in (a) were calculated.
}
  \label{fig:sinthetam}
\end{figure}

Figure \ref{fig:sinthetam} shows results of numerical simulations of transport
across a line of Gaussian potential wells.
These simulations were designed to model the experimental design of Ref.~\cite{ladavac04},
in which colloidal spheres are driven by flowing fluid past an inclined line of discrete
optical traps.  The driving force for this system is $F_0 \approx \xi a u$, where $\xi$ is
viscous drag coefficient corrected for hydrodynamic coupling to walls, $a$ is the
radius, and $u$ is the flow speed.
The sample trajectories in Fig.~\ref{fig:sinthetam}(a) were calculated for
$w = 0.4 \, b$ and 
$\eta(a) = 9.7 \, \sigma^2/a^2$ at a fixed orientation of $\theta = 17.5^\circ$.
The demonstrate that spheres with radii larger than $a = 0.1\, b$ are locked in to
the array of traps under these conditions, while smaller spheres escape.
Even a comparatively short array can resolve differences in radius of just a few
percent, suggesting that nanometer-scale resolution should be attainable for
hundred-nanometer-scale spheres in practical optical implementations.

Figure \ref{fig:sinthetam}(b) shows how the marginally locked-in angle varies
with size for this array.
The lower dashed curve in Fig.~\ref{fig:sinthetam} is the prediction of Eq.~(\ref{eq:lgcriterion}).
Its very good agreement with simulation results in this parameter range confirms
that the limiting argument establishes a useful lower bound on the $\theta_m$.
These results therefore confirm 
that fractionation by a line of traps offers exceptional size selectivity
in an appropriate range of conditions.
Figure \ref{fig:sinthetam}(b) also demonstrates that the locked-in fraction
can be deflected to large angles,
contrary to assertions in previous reports \cite{macdonald03}.

While $b > \sigma$ ensures optical fractionation's exponential size selectivity,
other considerations provide a basis for optimizing the inter-trap separation.
The total lateral deflection for a captured particle in an $N$-trap
array is $(N-1) \, b \, \sin \theta$. 
The array's efficiency can be defined accordingly as the lateral deflection per
trap:
$\Delta(a,b) = b \sin \theta.$
Choosing $b = 2 \sigma(a)$ optimizes this efficiency
at $\Delta = (4/e) \, V_0/F_j$.
This result, however,
does not necessarily optimize sensitivity to particle size.

The sensitivity may be formulated as
\begin{equation}
  S(a,b) \equiv \frac{\partial \Delta(a,b)}{\partial a},
\end{equation}
and is optimized by setting
\begin{equation}
  \frac{\partial S(a,b)}{\partial b} = \frac{\partial^2 \Delta(a,b)}{\partial b \partial a} = 0.
\end{equation}
This yields an optimal separation somewhat larger than that for maximum deflection:
\begin{equation}
  \label{eq:b}
  \frac{b^2}{4 \sigma^2} = 1 + \chi(a) + \sqrt{ 3 + \chi^2(a)},
\end{equation}
with
\begin{equation}
  \chi(a) = \frac{1}{2}\,\left[1 -  
    \frac{\eta^\prime(a)}{\eta(a)}\,\frac{\sigma(a)}{\sigma^\prime(a)}\right].
\end{equation}
Although fractionation by a line of optical traps has been demonstrated in practice
\cite{ladavac04},
optimization based on these criteria has yet to be
implemented.

\section{Conclusions}

Periodic potential-energy landscapes have exceptional promise for sorting 
continuous streams of mesoscopic objects.
Whether an object becomes locked in to a symmetry-selected direction through
the landscape or instead follows the direction of the driving force can depend
sensitively on size.  This can be shown quite generally for the separable
potentials considered in Secs.~\ref{sec:sine} and \ref{sec:separable}.
More subtle landscapes, which involve coupled motions in two or more
dimensions, are more difficult to analyze.  Approximate arguments and simulations
show that a particular one of these, a line of Gaussian wells, 
offers both exponential size selectivity
and clean binary separations.
More sophisticated, non-separable, higher-dimensional landscapes, such as
two-dimensional arrays of optical traps \cite{korda02b}, optical
lattices \cite{macdonald03}, and microfabricated post arrays \cite{duke97},
can distribute continuous distributions of objects into
discrete fractions \cite{gopinathan04}.  The analysis in this case
is made far more difficult by the lack of closed-form solutions
to the equations of motion, even in the deterministic limit.

Randomization by thermal forces substantially degrades the
selectivity with which a one-dimensionally modulated landscape can
retain objects.
A related study demonstrates that thermal forcing restructures the
pattern locked-in states in a two-dimensional array of potential wells \cite{gopinathan04},
and eventually wipes them out as the array grows in size.
This contradicts the assertion \cite{macdonald03} that thermally assisted hopping
can lead to exponential size selectivity.
Fortunately, the sorting processes discussed here, as well as their generalizations,
can be driven into the deterministic limit by increasing the driving and trapping
forces.

Continuous, continuously tuned chromatographic size separations should
have many applications in biological research, drug discovery,
and purification of mesoscale materials.
This Article outlines the basic principles by which they work, and suggests
considerations for their optimization for particular applications.

This work was supported by the National Science Foundation under Grant Number
DBI-0233971 and Grant Number DMR-0304906.


\end{document}